\documentclass[%
superscriptaddress,
amsmath,amssymb,
aps,
floatfix,
]{revtex4-2}

\usepackage{xcolor}
\usepackage{graphicx}
\usepackage{bm}
\usepackage{hyperref}

\usepackage{amsfonts}
\begin{document}


\title{Deep Learning with Photonic Neural Cellular Automata}

\author{Gordon H.Y. Li}
\affiliation{Department of Applied Physics, California Institute of Technology, Pasadena, CA 91125, USA}
\author{Christian R. Leefmans}
\affiliation{Department of Applied Physics, California Institute of Technology, Pasadena, CA 91125, USA}
\author{James Williams}
\affiliation{Department of Electrical Engineering, California Institute of Technology, Pasadena, CA 91125, USA}
\author{Robert M. Gray}
\affiliation{Department of Electrical Engineering, California Institute of Technology, Pasadena, CA 91125, USA}
\author{Midya Parto}
\affiliation{Department of Electrical Engineering, California Institute of Technology, Pasadena, CA 91125, USA}
\author{Alireza Marandi}
\email{marandi@caltech.edu}
\affiliation{Department of Applied Physics, California Institute of Technology, Pasadena, CA 91125, USA}
\affiliation{Department of Electrical Engineering, California Institute of Technology, Pasadena, CA 91125, USA}


\begin{abstract}
\noindent \textbf{Abstract:} Rapid advancements in deep learning over the past decade have fueled an insatiable demand for efficient and scalable hardware. Photonics offers a promising solution by leveraging the unique properties of light. However, conventional neural network architectures, which typically require dense programmable connections, pose several practical challenges for photonic realizations. To overcome these limitations, we propose and experimentally demonstrate Photonic Neural Cellular Automata (PNCA) for photonic deep learning with sparse connectivity. PNCA harnesses the speed and interconnectivity of photonics, as well as the self-organizing nature of cellular automata through local interactions to achieve robust, reliable, and efficient processing. We utilize linear light interference and parametric nonlinear optics for all-optical computations in a time-multiplexed photonic network to experimentally perform self-organized image classification. We demonstrate binary classification of images in the fashion-MNIST dataset using as few as 3 programmable photonic parameters, achieving an experimental accuracy of $98.0\%$ with the ability to also recognize out-of-distribution data. The proposed PNCA approach can be adapted to a wide range of existing photonic hardware and provides a compelling alternative to conventional photonic neural networks by maximizing the advantages of light-based computing whilst mitigating their practical challenges. Our results showcase the potential of PNCA in advancing photonic deep learning and highlights a path for next-generation photonic computers.
\end{abstract}

\maketitle
Deep learning models have demonstrated remarkable capabilities in numerous domains, ranging from computer vision to natural language processing, scientific discovery, and generative art~\cite{lecun2015deep,vaswani2017attention,jumper2021highly,goodfellow2014generative}. However, as the complexity and scale of these models continue to surge, a critical challenge emerges: the need for efficient and scalable hardware solutions to handle the ever-increasing computational demands. For example, recent trends show that the compute requirements for deep learning models are doubling approximately every 5-6 months~\cite{sevilla2022compute}. This is far outpacing improvements in conventional digital electronic computers, which has spurred the use of application-specific hardware accelerators such as Graphics Processing Units and Tensor Processing Units~\cite{jouppi2017datacenter}. In this context, the convergence of deep learning with photonics has emerged as a promising frontier, poised to redefine the landscape of neural network computation. By leveraging the distinct characteristics of light, photonic hardware can unlock unprecedented processing speeds, parallelism, and energy efficiencies that surpass the capabilities of traditional electronic architectures~\cite{shastri2021photonics,wetzstein2020inference}. To enable this new paradigm of photonic deep learning, much of the focus so far has been on developing the fundamental devices needed for crucial neural network operations. Indeed, there have been impressive demonstrations of photonics for linear operations such as matrix multiplication and convolutions~\cite{shen2017deep,feldmann2021parallel,xu202111}, as well as nonlinear activation functions such as rectified linear unit~\cite{feldmann2019all,li2022all,ashtiani2022chip}. These photonic building blocks are now comparable to or surpass their electronic counterparts in certain important computing metrics.

However, there has been comparatively less attention devoted towards studying system-level architectures for photonic neural networks (PNNs). This is crucial since photonics and electronics operate in entirely different regimes~\cite{miller2010optical}. The computational advantages of photonic building blocks can quickly diminish when used to implement conventional neural network architectures that were optimized for digital electronics~\cite{sze2017efficient}. Advancing photonic deep learning towards end-to-end and scalable photonic systems requires properly considering neural network architectures that can benefit from implementation with specific photonic hardware. For example, one important hurdle is that photonic devices are typically analog and noisy, requiring low effective bit-resolution to operate efficiently~\cite{tait2022quantifying}. This is detrimental for conventional deep learning architectures such as Multi-layer Perceptrons (MLPs) and Convolutional Neural Networks (CNNs), which have so far been mainstays for PNNs, because they are inherently susceptible to noise and small perturbations~\cite{szegedy2013intriguing,goodfellow2014explaining}. We expect that this problem will become increasingly pronounced as PNNs grow beyond small-scale demonstrations. Moreover, MLPs and CNNs require densely-connected layers with large numbers of parameters, which are challenging to realize in typical photonic platforms and current demonstrations of PNNs contain relatively small numbers of programmable parameters. Finally, PNNs are usually operated with fixed weights that cannot be rapidly updated in real-time. This constraint makes it difficult for PNNs to efficiently implement the complex structures of modern deep learning models, and also poses reliability concerns when generalizing to out-of-distribution data. 

To overcome these apparent disparities between photonics capabilities and conventional neural network architectures, we propose and experimentally demonstrate a novel type of PNN based on Neural Cellular Automata (NCA)~\cite{mordvintsev2020growing}. Cellular automata (CA) are computational models composed of a lattice of cells with states that follow an update rule, which defines how the state of a cell evolves over time based on the states of its neighboring cells~\cite{wolfram1983statistical,li2023photonic}. Inspired by biological systems, the local interactions between cells governed by the update rule gives rise to complex phenomena and emergent patterns at the global-scale~\cite{gardner1970fantastic}. Unlike conventional human-designed update rules, NCA harness the complex dynamics of cellular automata by using modern deep learning techniques to learn the local update rules needed to perform specific tasks such as regenerating patterns~\cite{mordvintsev2020growing}, self-classifying images~\cite{randazzo2020self}, and texture generation~\cite{niklasson2021self}. Our Photonic Neural Cellular Automata (PNCA) combines the advantages of photonic hardware with NCA to achieve self-organized image classification. The PNCA leverages a completely different methodology for computer vision tasks compared to previous PNNs based on MLPs or CNNs. Crucially, this enables noise-robust and fault-tolerant processing, as well as convenient measures of uncertainty for identifying anomalies and out-of-distribution data. Furthermore, PNCA achieves parameter-efficient solutions since the photonic hardware can operate with fixed weights and only needs to encode the parameters for local update rules instead of global network weights. The proposed PNCA approach can be generalized to suit a wide variety of existing photonic hardware, which can potentially greatly increase the functionality of PNNs and addresses several important challenges facing photonic deep learning. 

\begin{figure}[h]
\includegraphics[width=\linewidth]{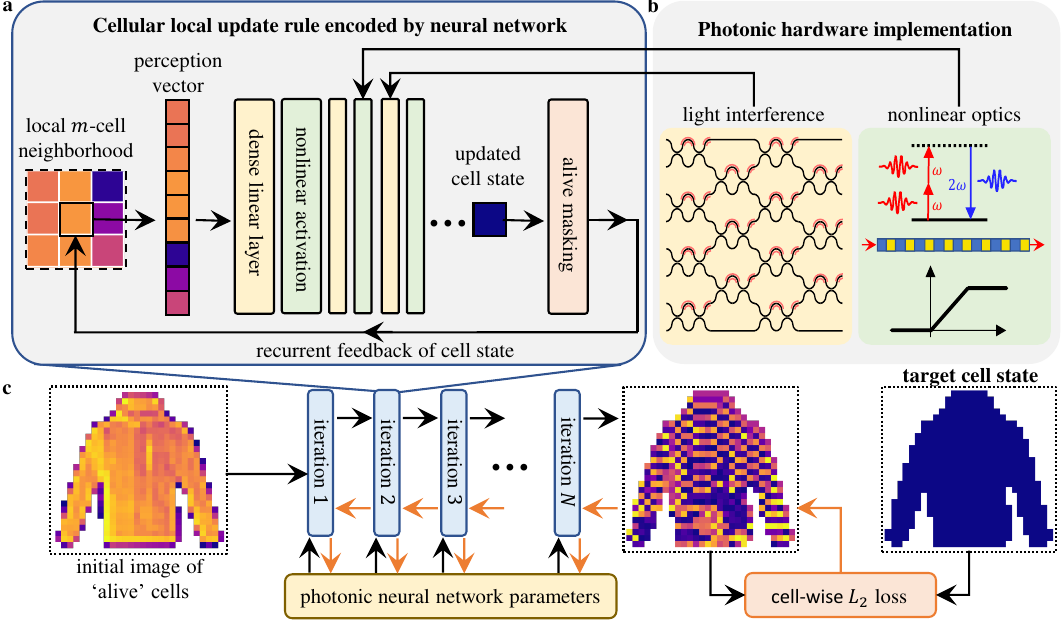}
\caption{\textbf{Photonic Neural Cellular Automata.} (a) A single iteration of PNCA consists of alive cells that are encoded into an optical signal, $m$ optical paths encoding a local $m$-cell neighborhood and perception vector for each cell, updating the state of each cell according to a local update rule represented by a neural network, and alive cell masking. (b) Photonic hardware encodes the local update rule, which includes linear operations implemented physically via light interference, and nonlinear operations implemented physically via nonlinear optics. (c) Backpropagation-through-time algorithm for training PNCA to learn a local update rule, which upon repeated iteration causes self-organization of cells for an image classification task. A cell-wise $L_{2}$ loss is used for optimizing the photonic neural network parameters.}
\label{fig:1}
\end{figure}

\section*{Results}
\subsection*{PNCA architecture}
The key concepts of the general PNCA architecture are shown in Fig.~\ref{fig:1}, which can be adapted to suit a wide range of different photonic hardware platforms. For computer vision tasks, each pixel in the input image corresponds to a cell in the PNCA. Cells are designated as either \textit{alive} or \textit{dead} through an alive masking procedure. This can be done by setting a threshold for the initial pixel value, below which the cell is considered dead. Only alive cells are actively updated by the PNCA, whereas dead cells can influence the updates of alive cells but are otherwise quiescent. The cell state updates according to a rule that depends on the cells in a local $m$-cell neighborhood. For example, Fig.~\ref{fig:1}a shows the prototypical Moore neighborhood composed of the cell and the 8 cells that surround it. Other types of local cell neighborhoods are also possible. In the PNCA, the optical field corresponding to each cell is split into $m$ optical paths to define the desired $m$-cell neighborhood for the local update rule. The local update rule for the PNCA is encoded by the photonic hardware, which accepts the $m$ inputs given by the $m$-cell neighborhood and outputs the next cell state. Although Fig.~\ref{fig:1}a only shows each cell state having a single channel, this can also be extended to multiple channels (e.g. RGB color image channels) by increasing the inputs and outputs accordingly. In general, the programmable photonic hardware contains feed-forward layers with linear operations which can be implemented through meshes of Mach-Zehnder interferometers~\cite{shen2017deep}, photonic cross-bar arrays~\cite{feldmann2021parallel}, micro-ring resonator weight banks~\cite{tait2017neuromorphic}, or other linear photonic devices~\cite{ashtiani2022chip,xu202111}. In addition, there must also be layers performing nonlinear activations such as photonic devices based on optoelectronic measurement-feedback~\cite{ashtiani2022chip,williamson2019reprogrammable} or nonlinear-optical crystals~\cite{li2022all,feldmann2019all}. This kind of feed-forward programmable photonic hardware specifying a single input-output function has been used in previous PNNs. However, for PNCA, the key difference is that the photonic hardware only needs sparse connections and enough parameters to encode for the local update rule as shown in Fig.~\ref{fig:1}b, which is usually orders-of-magnitude fewer than the number of parameters needed to encode global network weights in fully-connected layers for MLPs or CNNs. In other words, the parameter-efficient PNCA architecture can enable existing PNN hardwares with relatively few parameters to perform larger and more complicated tasks than otherwise possible in conventional neural network architectures. Furthermore, this local update rule can more easily tolerate the use of fixed-weights after training since every cell follows the same update rule. Finally, the output is recurrently fed back to update the cell state for the next iteration. This can be accomplished by photodetection and electro-optic feedback or by using all-optical feedback lines. 

Unlike conventional CA with discrete cell states~\cite{wolfram1983statistical}, NCA use cell states that are continuous-valued~\cite{mordvintsev2020growing}, which allows the model to be end-to-end differentiable and compatible with gradient-descent based learning algorithms. In this work, we consider the task of self-organized image classification. The target output after the final iteration is to have every alive cell in the state that corresponds to the class label for the input image. The alive cells must form this collective agreement through only the local interactions defined by repeated iteration of the update rule. This can be interpreted as a kind of recurrent neural network, which can be trained using the standard backpropagation-though-time algorithm~\cite{werbos1990backpropagation} as shown in Fig.~\ref{fig:1}c. Using a cell-wise $L_{2}$ loss was found to give better performance compared to cross-entropy loss of labels, which is more commonly used for image classification tasks~\cite{mordvintsev2020growing}. The training can either be done \textit{in situ} by performing the forward pass in PNCA to more accurately capture the physics, or completely digitally by simulating the photonic hardware with noise~\cite{wright2022deep,pai2023experimentally}. 

\begin{figure}[t]
\includegraphics[width=\linewidth]{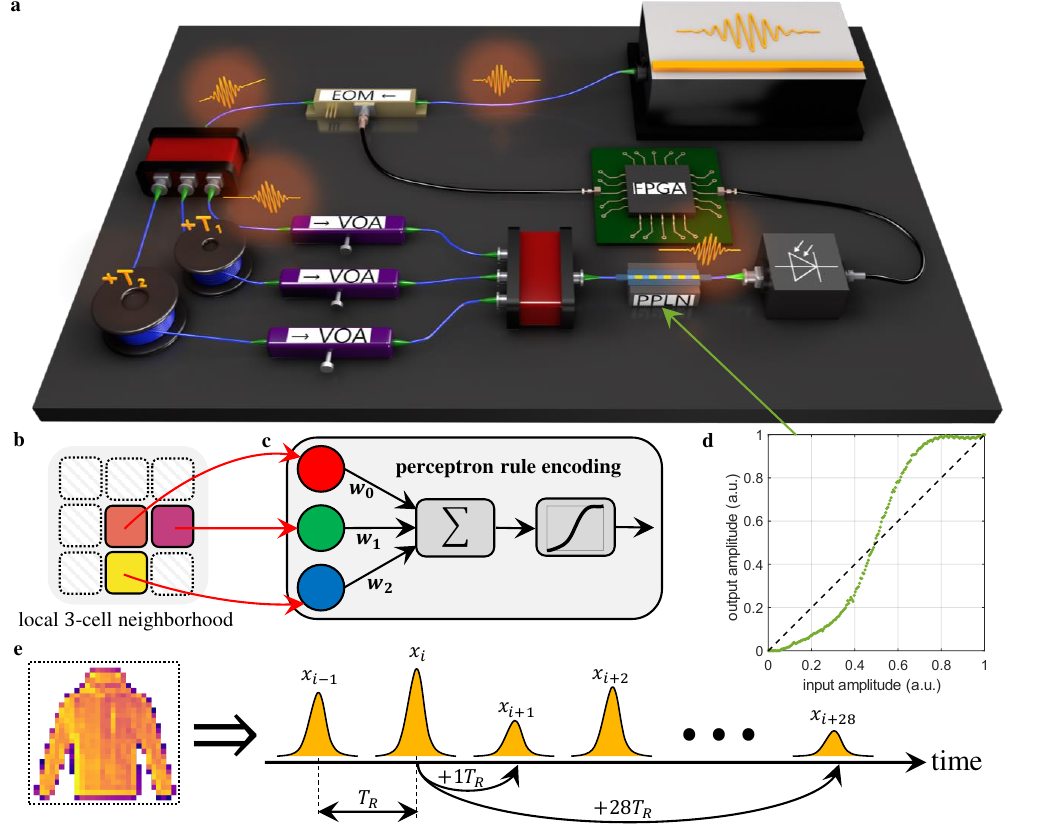}
\caption{\textbf{Experimental setup for PNCA.} (a) Schematic of the experimental setup. Pulses of light produced by a mode-locked laser pass through an electro-optic modulator (EOM) and are split into optical fiber delay lines (blue lines) with relative delays $T_{1}$ and $T_{2}$. Linear dot product weights are programmed by tuning the variable optical attenuator (VOA) in each delay line. Nonlinear activation using a periodically-poled lithium niobate (PPLN) waveguide is performed following the coherent
interference of light pulses, with the resultant amplitudes stored on a field-programmable gate array (FPGA) and reinjected (black lines) to drive the
input EOM for the next iteration. (b) Local 3-cell neighborhood enforced by relative delays $T_{1}$ and $T_{2}$. (c) The local update rule is encoded by a single perceptron with 3 programmable parameters. (d) PPLN nonlinear activation function. (e) Cells representing pixels of an image are encoded by the amplitude of light pulses with repetition period $T_{R}$ in a synthetic temporal dimension. For example, pulses can be coupled using optical delay lines with $T_{1}=+1T_{R}$ and $T_{2}=+28T_{R}$ to implement the local 3-cell neighborhood shown in (b) for fashion-MNIST images.}
\label{fig:2}
\end{figure}

\subsection*{Experimental realization of PNCA}
We used a time-multiplexed scheme and commercially-available optical-fiber components to experimentally demonstrate proof-of-concept for a simple version of PNCA as shown in Fig.~\ref{fig:2}. Each cell state is given by the amplitude of a laser light pulse generated by a mode-locked laser with a fixed repetition rate such that the cells are inputted one at a time in a flattened 1D lattice by raster scanning across the 2D image. In this way, each cell occupies a time-bin site in a synthetic temporal dimension~\cite{leefmans2022topological}. Therefore, distances in a real-space lattice correspond to time-differences in the temporal dimension and cells at different lattice sites can be made to interact by using temporal delay lines. The pulse amplitude/phase representing the cell state is set using an electro-optic modulator (EOM), and the pulse is then split between 3 temporal optical delay lines with relative delays $T_{1}$ and $T_{2}$ chosen to enforce the desired 3-cell local neighborhood shown in Fig.~\ref{fig:2}b. In this simple example, the local update rule is encoded by a single perceptron neuron shown in Fig.~\ref{fig:2}c, which consists of a linear dot product followed by a nonlinear activation function. The dot product is achieved by coherent interference of the optical delay lines, each equipped with a variable optical attenuator (VOA) to program the desired weights. The nonlinear activation is performed using depleted second harmonic generation in a reverse-proton exchange periodically-poled lithium niobate waveguide~\cite{langrock2007fiber}. This produces a sigmoid-like function as shown in Fig.~\ref{fig:2}d. Thus, the computations in the local update rule are achieved all-optically. Overall, the local update rule contains only 3 programmable parameters, but can still perform complex tasks. Finally, the cell state is measured using a photodetector, stored on a field-programmable gate array (FPGA), and electro-optically re-injected for the next iteration after alive-cell masking. 

\begin{figure}[t]
\includegraphics[width=\linewidth]{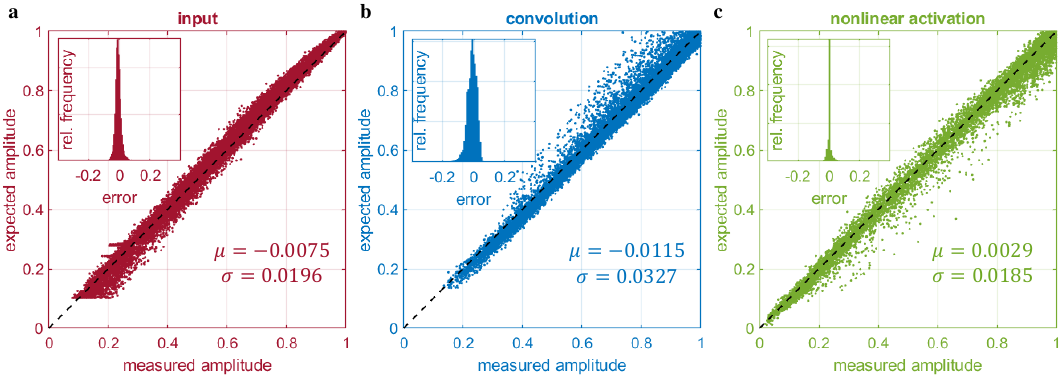}
\caption{\textbf{Measurements of noise and errors in PNCA operations.} Expected vs. measured light amplitude for (a) input cell state by EOM, (b) linear dot product by coherent intereference and (c) nonlinear activation by PPLN. Each scatter point represents an alive cell from the 200 images tested. The top right insets show the histograms for the error (expected amplitude $-$ measured amplitude) in each case and the bottom right shows the mean and standard deviation, respectively.}
\label{fig:3}
\end{figure}

A crucial aspect of photonic hardware is that it is analog and noisy. A key advantage of the PNCA architecture is that it is fault-tolerant and robust to noise due to the self-organizing nature of the cell states. We rigorously characterized the noise and errors in our PNCA implementation, which arises from three main operations: (1) the input cell state due to thermal and electronic noise in the EOM, (2) the linear dot product due to phase noise and imperfect pulse temporal overlap in the coherent interference, and (3) the nonlinear activation due to thermal noise and photorefractive effects in the PPLN. We characterized these errors using 200 test images. The expected vs. measured amplitudes of alive cells in these images are shown in Fig.~\ref{fig:3}. The mean and standard deviation of the errors (expected amplitude $-$ measured amplitude) achieved in our system are typical of photonic hardware, and we show that this is tolerable for the PNCA architecture due to its noise-robustness.

\subsection*{Self-organized image classification}
\begin{figure}[t]
\includegraphics[width=\linewidth]{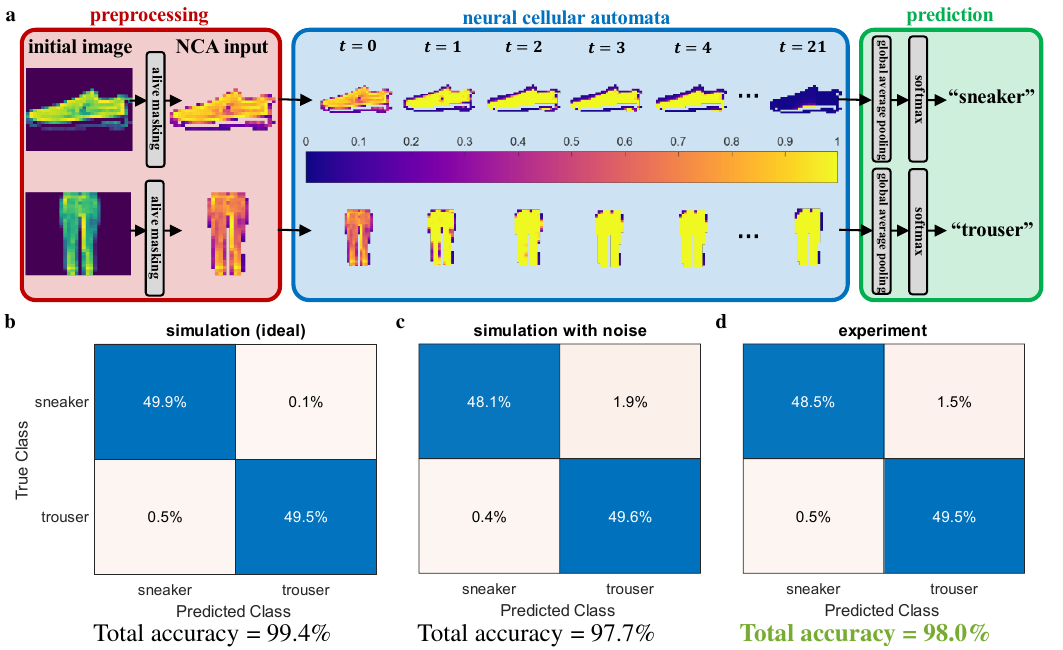}
\caption{\textbf{Experimental results for fashion-MNIST binary image classification.} (a) Information flow for the PNCA trained to classify images of sneakers and trousers, beginning with alive cell masking, followed by $t=21$ iterations of the trained PNCA. The predicted image label is obtained by global average pooling and softmax classification of the final self-organized alive cells. Confusion matrices for (b) idealized simulation model, (c) noisy simulation model, and (d) experiment.}
\label{fig:4}
\end{figure}

We trained the experimental PNCA to perform binary image classification using the fashion-MNIST dataset consisting of $28\times28$ pixel gray-scale images of clothing items~\cite{xiao2017fashion}. For example, Fig.~\ref{fig:4}a shows how the PNCA can classify images of sneakers and trousers. The alive cell masking is performed by designating any pixel with initial value $\alpha>0.1$ as an alive cell, and all other pixels as dead cells with constant value of zero. Each input image was iterated for $t=21$ time steps in the PNCA, which was sufficient for the cells to reach an approximate global agreement. The alive cells self-organize to have state values close to zero (unity) for images of sneakers (trousers). Finally, the predicted image label is obtained in postprocessing by performing global average pooling of the final alive cell states followed by softmax classification. In this case, a global average closer to zero (unity) indicates that the predicted image label is sneaker (trouser).

The training procedure was performed digitally using an idealized simulation model of the PNCA that had no noise. The confusion matrix for the idealized model is shown in Fig.~\ref{fig:4}b, which yielded a final test accuracy of $99.4\%$. Next, the trained model parameters were frozen, and the model was tested again but with additional simulated Gaussian noise for each operation, matching the noise characteristics shown in Fig.~\ref{fig:3}. The confusion matrix for the noisy model is shown in Fig.~\ref{fig:4}c, which has a slightly lower final test accuracy of $97.7\%$. The trained model parameters were implemented in the experimental PNCA by appropriately tuning the VOAs. The confusion matrix for the experimental result is shown in Fig.~\ref{fig:4}d and has a final test accuracy of $98.0\%$. This experimental test accuracy is in close agreement with the simulated noisy model, which shows that the PNCA operates as desired and can successfully tolerate the use of noisy photonic hardware. No special training or noise regularization techniques were used for the PNCA. We emphasize that the robustness emerges through the local interactions between cells forming a global agreement. Therefore, even if one cell fails, the collective state can still persist. This is in contrast to conventional neural network architectures such as MLPs and CNNs, which are highly susceptible to noise and adversarial attacks~\cite{szegedy2013intriguing}. 

\subsection*{Out-of-distribution data}
\begin{figure}[b]
\includegraphics[width=\linewidth]{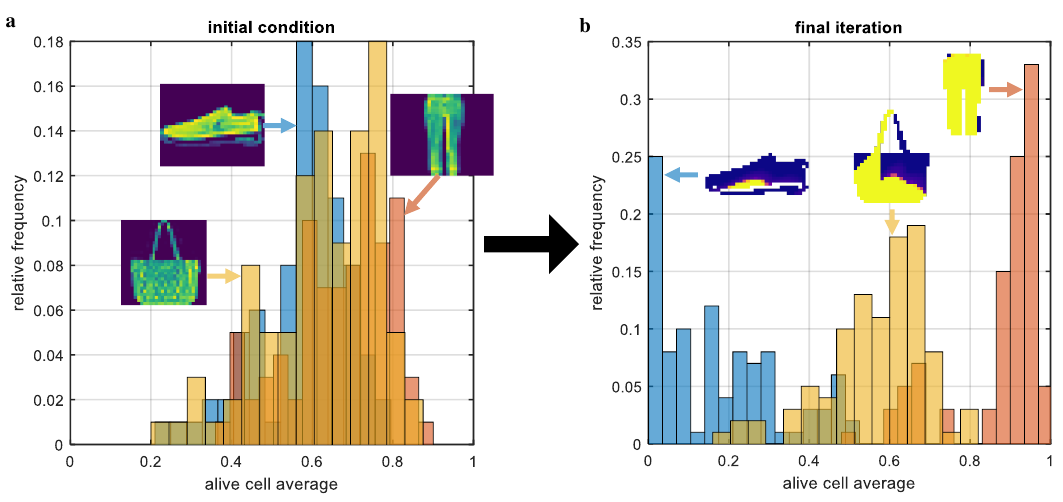}
\caption{\textbf{Recognizing out-of-distribution data.} Histograms of alive cell averages for (a) initial condition and (b) final iteration of test images of sneakers (blue), trousers (red), and out-of-distribution bags (yellow).}
\label{fig:5}
\end{figure}
Furthermore, conventional neural networks are prone to making overconfident predictions and failing to generalize to out-of-distribution data~\cite{guo2017calibration}. This lack of reliability is especially problematic for photonic deep learning in which the weights are fixed and online learning is not practical. The NCA approach addresses this shortcoming by using the average state value of all alive cells as a built-in measure of uncertainty. We experimentally demonstrated this for PNCA by using the same network as before that was trained on images of sneakers and trousers. Now, we test the PNCA on images of bags, which is an out-of-distribution class that the PNCA was not exposed to during training. The distributions for the alive cell averages of the sneaker, trouser, and bag classes are shown for the initial test images in Fig.~\ref{fig:5}a. It clearly shows that the initial distributions for alive cell averages closely overlap between all classes. Upon iteration of the local update rule that was learned during training, the PNCA is able to successfully separate the distributions for sneaker and trouser, with final alive cell averages of 0.1743 and 0.8742, respectively, as shown in Fig.~\ref{fig:5}b. In this case, the difference between the final alive cell average and zero/one indicates the uncertainty in the prediction. However, the final alive cell average for out-of-distribution test images of bags is 0.5682, which is close to 0.5 and means that the cells did not reach a global agreement. This shows that the PNCA can use the alive cell average as a proxy for uncertainty and to detect out-of-distribution data. Unlike for conventional neural network architectures, neither special training/inference techniques nor additional training data is required. 

\section*{Discussion}
In summary, we have proposed and experimentally demonstrated a novel approach to photonic deep learning based on PNCA. It addresses several system-level challenges in previous photonic neural networks, and can serve as a general architecture for a wide variety of photonic hardware platforms. In particular, we showed that PNCA enables noise-robust fault-tolerant image classification through local interactions between cells with an inherent measure of uncertainty based on alive cell averages. Moreover, the efficient PNCA model encoding requires orders of magnitude fewer parameters compared to MLPs or CNNs. Our single perceptron neuron rule encoding can be straightforwardly extended to a shallow neural network with a greater number of programmable parameters to perform more complicated and larger-scale computer vision tasks. For example, we focused on binary image classication for simplicity, but it is possible to perform image classification with more classes if the number of output neuron channels is increased. Furthermore, we only used standard backpropagation training and did not employ any special training or regularization techniques. More advanced noise-aware or physics-aware training schemes~\cite{wright2022deep} are also compatible with the PNCA architecture and may further increase performance. We used a time-multiplexed photonic network based on a synthetic temporal dimension, however, it is also possible to use an analogous PNCA approach based on other synthetic dimensions such as frequency dimensions~\cite{yuan2018synthetic}. Our work therefore highlights a clear path to advancing photonic deep learning based on PNCA and paves the way for next-generation photonic computers.
\bibliography{apssamp}
\newpage
\section*{Methods}
\subsection*{Experimental Setup}
A more detailed schematic of the experimental setup is shown in Extended Data Fig.~\ref{fig:S1}. A femtosecond laser source (MenloSystems FC1500-250-WG) produces pulses of light at a fixed repetition rate of $\sim250~\mathrm{MHz}$. The light pulses are filtered using a $200~\mathrm{GHz}$ band-pass filter with center wavelength $\sim1550~\mathrm{nm}$ to stretch the pulse length to $\sim5~\mathrm{ps}$ and reduce the effects of dispersion. The light pulses are photodetected (MenloSystems FPD610-FC-NIR) as a reference clock signal for the FPGA (Xilinx Zynq UltraScale+ RFSoC) to eliminate timing drift between the optical and electronic signals. The FPGA drives an electro-optic modulator (IXblue MXAN-LN-10) that is used to modulate the amplitude of the light pulses. The light pulses are split into a 3-path interferometer by cascading 50:50 optical fiber splitters. Two paths of the interferometer have delays $+1T_{R}$ and $+28T_{R}$, respectively, relative to the shortest path, where $T_{R}$ is the repetition period of the light pulses. The relative delays in each arm are set using a combination of optical fiber patch cords and free-space delay stages. Tuning the free-space coupling efficiency also acts a variable optical attenuator to set the relative amplitude weight in each arm. The output of the 3-arm interferometer is tapped using a 90:10 optical fiber splitter. The $10\%$ tap is photodetected (Newport New Focus Model 2053) and used as an electronic locking signal input to a proportional-integral derivative controller (Red Pitaya). The electronic locking signal output is amplified (Thorlabs Piezo Controller MDT693B) and drives fiber phase-shifters (General Photonics FPS-002-L) that stabilize the relative phases of each delay arm. The $90\%$ output of the 3-arm interferometer is amplified using an erbium-doped fiber amplifier (Thorlabs Fiber Amplifier 1550nm PM) and filtered using a $200~\mathrm{GHz}$ band-pass filter to reduce the amplified spontaneous emission noise. The amplified light pulses pass through a $40~\mathrm{mm}$ long reverse-proton exchange PPLN waveguide~\cite{langrock2007fiber} that is heated to $\sim52^{\circ}\mathrm{C}$ with a thermocouple controller. The PPLN waveguide contains a wavelength division multiplexer on the output to separate the fundamental harmonic centered at $\sim1550~\mathrm{nm}$ and the second harmonic centered at $\sim775~\mathrm{nm}$. The second harmonic output is dumped and the fundamental harmonic is photodetected (Thorlabs DET08CFC). The final photodetected signal is read as a time trace using an oscilloscope (Tektronix MSO6B) and light pulse amplitude values are stored on the FPGA to be electro-optically reinjected. All optical fiber paths are single-mode polarization-maintaining (PM). 
\subsection*{Photonic Neural Cellular Automata Model}
The Neural Cellular Automata (NCA) comprises a lattice of cells indexed by lattice site numbers $i\in\mathbb{N}$ with states $\mathbf{x_{i}}\in\mathbb{C}^{d}$, where $d$ is the number of channels for each cell.  Each cell interacts locally in an $m$-cell neighborhood $\mathbb{M}_{i}$ according to a fixed update rule. We consider discrete-time synchronous updates $t\in\mathbb{N}$ for cells:
\begin{equation}
    \mathbf{x_{i}}(t+1)=f_{\theta}(\mathbf{x_{m_{i1}}}(t),\mathbf{x_{m_{i2}}}(t),\mathbf{x_{m_{i3}}}(t),\cdots)\ ,
    \label{eq:1}
\end{equation}
where $m_{i1},m_{i2},m_{i3},\cdots\in\mathbb{M}_{i}$ are the lattice sites in the local neighborhood of the $i^{\mathrm{th}}$ cell and $f_{\theta}:\left(\mathbb{C}^{d}\right)^{m}\rightarrow\mathbb{C}^{d}$ is the local update rule. The local update rule is parameterized by $\{\theta\}$ and is differentiable so that it can be trained using modern deep learning techniques. For example, $f_{\theta}$ can represent a neural network. The key aspect is that the update rule $f_{\theta}$ is the same for all cells and all time steps.  

We experimentally demonstrated a simple version of NCA implemented directly on analog photonic hardware, which we call Photonic Neural Cellular Automata (PNCA). In PNCA, lattice sites are represented by laser light pulses in time bins of a synthetic temporal dimension with a fixed repetition period $T_{R}$ and cell states are represented by the complex amplitude of the light pulses. For simplicity, we consider a single image channel $d=1$ and the local update rule $f_{\theta}$ encoded by a single perceptron neuron with an $m=3$ neighborhood as shown in Fig.~\ref{fig:2}b,c. The temporal delay lines $T_{1}=+1T_{R}$ and $T_{2}=+28T_{R}$ set the desired local cell neighborhood and the VOAs in each arm of the 3-arm interferometer set the desired weights $\{w_{0},w_{1},w_{2}\}\in[-1,+1]$. The PIDs are used to enforce a relative phase of $0$ for constructive interference, or conversely a relative phase of $\pi$ for destructive interference. Therefore, at the output of the 3-arm interferometer, the combined result of the delay lines, VOAs, and phases can be summarized as a linear dot product or sliding convolutional filter:
\begin{equation}
y_{i}(t)=w_{0}x_{i}(t)+w_{1}x_{i+T_{1}}(t)+w_{2}x_{i+T_{2}}(t)\ ,
\end{equation}
where the result of the linear operation $y_{i}(t)$ is fed into a PPLN to perform a nonlinear activation function:
\begin{equation}
    x_{i}(t+1)=g(y_{i}(t))\ ,
\end{equation}
where $g$ is the sigmoid-like function shown in Fig.~\ref{fig:2}d. The PNCA approach is very general and Eq.~\ref{eq:1} can be implemented using more complicated photonic hardware platforms with different cells neighborhoods, more neurons, deeper layers, and more programamble parameters. 
\subsection*{Experimental Procedure}
The input modulator was calibrated by using a sequence of 200 consecutive light pulses and performing a linear voltage sweep of the input EOM, which was DC biased open. The peak pulse amplitude or maximum value in each time bin (i.e. pulse repetition period) of the measured time trace was used to construct a look-up table for the voltage-to-light amplitude conversion. To input a specific $28\times28$ fashion-MNIST image, the 2D pixel map was unrolled column-wise to form a $784\times1$ vector of input cell values. Alive masking was applied such that any initial pixel value $<0.1$ was designated as a dead cell. The accuracy of the input operation was checked by measuring the difference between the measured input cell states and the expected value, such as shown in Extended Data Fig.~\ref{fig:S2}. The aggregate results are shown in Fig.~\ref{fig:2}a. The desired weights for the linear dot products were set by tuning the coupling efficiency of the free-space delay stages in each temporal delay arm of the 3-arm interferometer. The power was directly measured in each arm to roughly tune the attenuation factor, and then fine-tuning of the weight was performed by checking the result of the linear interference matched the expected value like in Extended Data Fig.~\ref{fig:S2}. A standard Pound-Drever-Hall locking scheme was used to stabilize the relative phases in each delay arm to either $0$ or $\pi$ to ensure coherent interference. It is also possible to make use of the full complex amplitude of light, although we restricted our attention to only real values. The relative delays in each temporal delay line was set roughly using optical fiber patch cords, then fine-tuned using free-space delay stages to ensure maximal temporal overlap between interfering light pulses. The aggregate results of the linear dot product or convolution operation are shown in Fig.~\ref{fig:2}b. The temperature of the PPLN was fine-tuned around $52^{\circ}\mathrm{C}$ until maximal average power was measured on the output second-harmonic given a small input fundamental harmonic average power $\sim1~\mathrm{mW}$. The PPLN nonlinear activation function shown in Fig.~\ref{fig:2}d was measured using a sequence of consecutive light pulses with linearly increasing input amplitude. To ensure stable operation over long-periods of time ($>12$ hours) throughout the experiment, we regularly check that the calibrated PPLN nonlinear activation function remains the same and does not change significantly due to photo-refractive or thermal effects. The measured values for PPLN nonlinear activations were also compared against the expected simulated values as shown in Extended Data Fig.~\ref{fig:S2}. The aggregate results of the PPLN nonlinear activation operation are shown in Fig.~\ref{fig:2}c. To perform self-organized image classification using the experimental PNCA, the input modulator was first calibrated. Then, the PPLN nonlinear activation function was measured, and a simulated digital model of the PNCA was trained (see Model Training) to determine the optimal weights to be set in the temporal delay lines. The light pulse amplitudes were stored digitally on the FPGA in between iterations, however, the iteration feedback can also be performed all-optically using an optical fiber cavity. 

\subsection*{Model Training}
The PNCA can be trained using the standard backpropagation-through-time algorithm for recurrent neural networks if a differentiable model of the update rule $f_{\theta}$ is known. The goal is to learn the parameters $\{\theta\}$ for a particular task such as self-organized image classification. We consider a cell-wise $L_{2}$ loss at each time step:
\begin{equation}
    L=\frac{1}{T}\sum_{t=1}^{T}\frac{1}{N}\sum_{i=1}^{N}||\mathbf{x_{i}}(t)-\mathbf{z_{i}}||^{2}\ ,
\end{equation}
where $\mathbf{z_{i}}$ is the target state for the $i^{\mathrm{th}}$ cell. The parameter values are updated using stochastic gradient descent:
\begin{equation}
    \theta^{[l+1]}=\theta^{[l]}-\alpha\nabla L(\theta^{[l]})\ ,
\end{equation}
where $l$ is the epoch number and $\alpha>0$ is the learning rate. The gradient $\nabla L$ is calculated by unrolling the network in time for $T$ time steps and applying the chain rule or automatic differentiation. More complicated gradient-based optimization such as stochastic gradient descent with momentum or adaptive moment estimation can also be used to perform parameter updates. We trained a PNCA to perform binary image classification of sneakers and trousers classes from the fashion-MNIST dataset using $5000$ training and $420$ validation images for each class, learning rate of $\alpha=0.002$, and $200$ training epochs. An example of a training curve is shown in Extended Data Fig.~\ref{fig:S3}.

\subsection*{Data Availability}
The data used to generate the plots and results in this paper are available from the corresponding author upon reasonable request.
\subsection*{Code Availablility}
The code used to analyze the data and generate the plots for this paper is available from
the corresponding author upon reasonable request.
\subsection*{Acknowledgments}
The authors acknowledge support from ARO grant no. W911NF-23-1-0048, NSF grant no. 1846273 and 1918549, Center for Sensing to Intelligence at Caltech, and NASA/JPL. The authors thank NTT Research for their financial and technical support. G.H.Y.L acknowledges support from the Quad Fellowship. 
\subsection*{Author Contributions}
All authors contributed to this manuscript.
\subsection*{Competing Interests}
The authors declare no competing interests.

\newpage
\setcounter{figure}{0}
\renewcommand{\figurename}{Extended Data Fig.}
\begin{figure}[h]
\includegraphics[width=\linewidth]{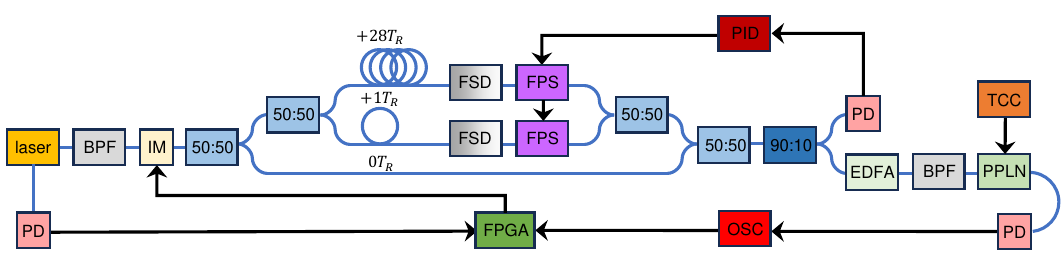}
\caption{\textbf{Detailed schematic of experimental setup.} BPF: band-pass filter, IM: intensity modulator, FSD: free-space delay stage, FPS: fiber phase-shifter, EDFA: erbium-doped fiber amplifier, PD: photodetector, PPLN: periodically-poled lithium niobate, TCC: thermocouple controller, PID: proportional integral derivative controller, OSC: oscilloscope, FPGA: field programmable gate array. Blue lines represent single-mode polarization-maintaining optical fiber paths and black lines represent electronic connections.}
\label{fig:S1}
\end{figure}
\newpage
\begin{figure}[h]
\includegraphics[width=\linewidth]{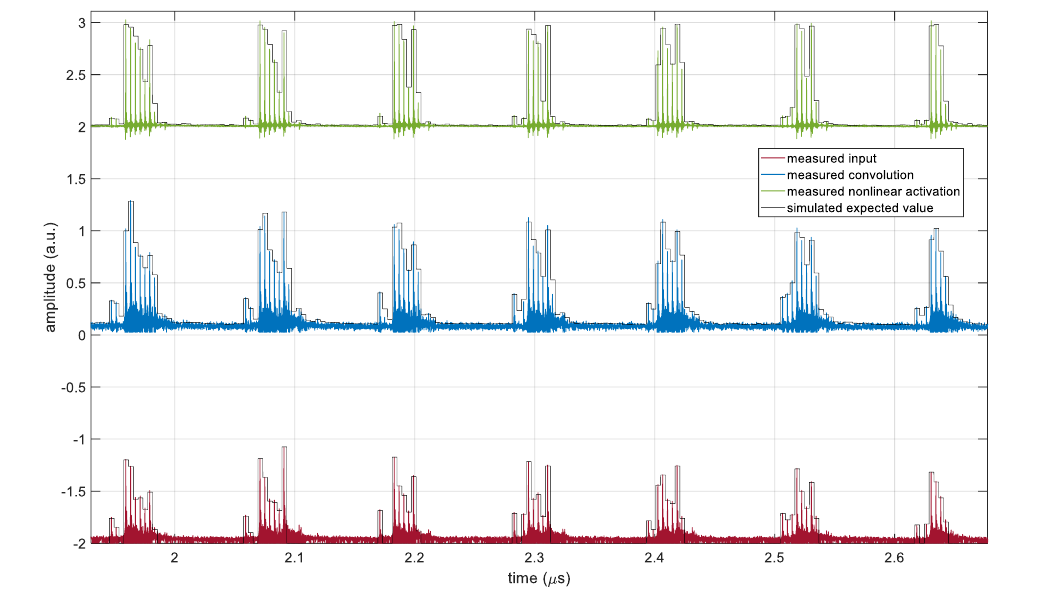}
\caption{\textbf{Example of experimentally-obtained time traces in photonic neural cellular automata.} Oscilloscope time traces for a portion of one iteration of one image in the photonic neural cellular automata showing the results of the input cell states (red line), linear convolution (blue line), and nonlinear activation (green line) compared against the expected values based on digital simulations (black lines). The peak or maximum amplitudes of each light pulse, which occupy a specific time bin in the synthetic temporal lattice, agrees well with the expected cell state values and shows that the calibration was accurate. The time traces for each operation are shifted vertically and corresponding cell lattice sites are aligned horizontally in time for clarity.}
\label{fig:S2}
\end{figure}
\newpage
\begin{figure}[h]
\includegraphics[width=\linewidth]{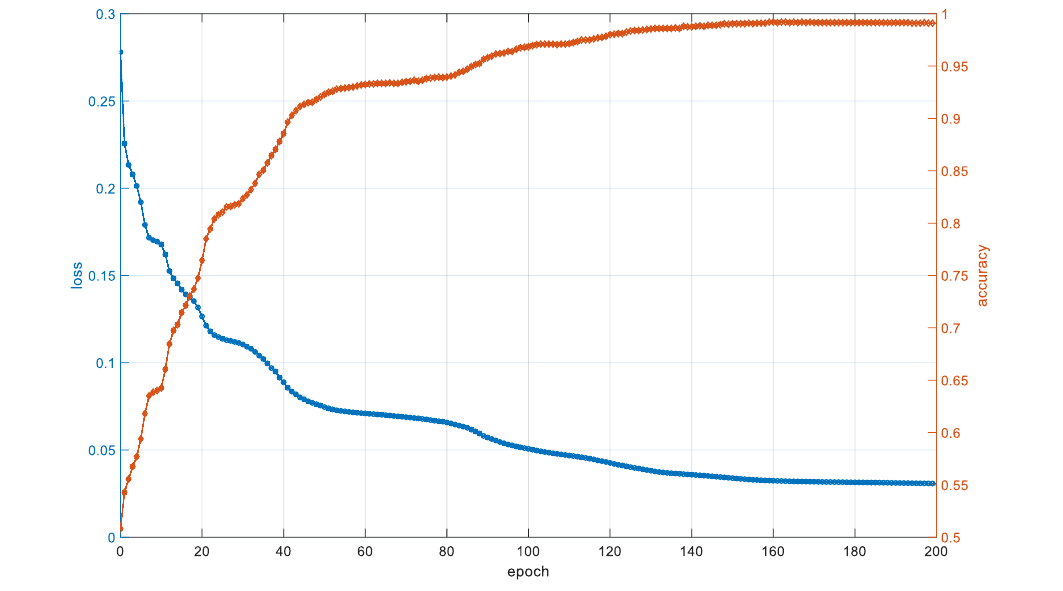}
\caption{\textbf{Training progress for photonic neural cellular automata.} Cell-wise $L_{2}$ loss (blue line) and classification accuracy (orange line) for training PNCA to perform binary image classification of sneakers and trousers classes from fashion-MNIST dataset.}
\label{fig:S3}
\end{figure}
\end{document}